\documentclass[12pt]{article}
\usepackage{amsthm,amsmath,latexsym,amssymb,amsfonts,amscd}
\usepackage{graphics,lscape,fancyhdr,array,stmaryrd,euscript}
\usepackage{ifthen,empheq}
\usepackage{sidecap}
\usepackage{ifpdf}
\usepackage{verbatim}
\usepackage{color}
\usepackage{relsize}
\numberwithin{equation}{section}
\usepackage{hyperref}
\usepackage[numbers,sort&compress]{natbib}
\usepackage[nottoc]{tocbibind}

\newcommand{\w}[1]{\widetilde{#1}}
\renewcommand{\r}{\rho}
\newcommand{\p}{\partial}

\begin{document}

\hfill
\vskip 0.01\textheight
\begin{center}
{\Large\bfseries 
Poisson Gauge Theories in Three Dimensions: \\[5mm]Exact Solutions and Conservation Laws}
\vskip 0.03\textheight
\renewcommand{\thefootnote}{\fnsymbol{footnote}}
Alexey \textsc{Sharapov}
\& David \textsc{Shcherbatov}
\renewcommand{\thefootnote}{\arabic{footnote}}
\vskip 0.03\textheight
{\em Physics Faculty, Tomsk State University, \\Lenin ave. 36, Tomsk 634050, Russia}\\
\vspace*{5pt}

\vskip 0.05\textheight

\begin{abstract}
We investigate Maxwell--Chern--Simons theory on a three-dimensional noncommutative spacetime endowed with a constant spacelike Poisson structure. By exploiting the residual rotational symmetry, we construct exact classical solutions corresponding to pointlike electric and magnetic charges. We demonstrate that noncommutativity acts as a natural regulator, ensuring a finite total electromagnetic energy and thereby resolving the classical self-energy divergence. Furthermore, some of these solutions exhibit a non-perturbative dependence on the noncommutativity parameter and allow for the generation of an arbitrary magnetic flux. We also present a noncommutative generalization of Gauss's law, providing a robust framework for the physical interpretation of these exact solutions.
\end{abstract}

\end{center}

\section{Introduction}
Noncommutative gauge theories provide a robust framework for exploring quantum field theory on noncommutative spacetime manifolds \cite{Douglas:2001ba, Szabo:2001kg, hersent2023gauge}. The introduction of a noncommutative structure inherently leads to non-locality and non-linearity in the field equations, which significantly complicates the search for exact solutions. Nevertheless, obtaining explicit solutions is essential for a consistent physical interpretation, as they reveal the field dynamics in the noncommutative regime and facilitate a direct comparison with their commutative counterparts.

 Normally, all vertex, monopole, or instanton solutions known in conventional gauge theory extend to the noncommutative setting, with singularities (if present) smeared out due to uncertainty in position space (see \cite{Douglas:2001ba} for a review). While these solutions are indispensable for the non-perturbative analysis of quantum fields, they do not always represent the primary focus of low-energy physics, which is centered on wave propagation and fields generated by point sources. Despite intensive research over the past two decades, Coulomb-like and wave-like solutions in noncommutative backgrounds remain largely unexplored. For various perturbative studies in this direction, we refer to \cite{cai2001superluminal, abe2003duality, mariz2007dispersion, gaete2004coulomb, stern2008noncommutative, stern2008particlelike, adorno2011noncommutative, abla2023effects}.

Given that spacetime noncommutativity must be sufficiently small to remain consistent with current observational data, it is natural to employ a semiclassical approach known as \textit{Poisson gauge theory} \cite{kupriyanov2020non,Kupriyanov:2020sgx, kupriyanov2021poisson, Kupriyanov_2021, kupriyanov2024symplectic, sharapov2024poisson}. In this framework, the noncommutative structure is encoded via a Poisson bracket on the spacetime manifold. Poisson gauge theory is expected to be valid at scales much larger than the characteristic length scale, yet it remains self-consistent with respect to gauge symmetry and is governed by local equations of motion.

Despite these advantages, the inherent non-linearity of Poisson gauge theories presents a formidable challenge. Unlike commutative theories, which benefit from the principle of superposition and the machinery of Green's functions, the non-linear structure of Poisson gauge theory leads to intricate field behaviors that are often analytically opaque. Consequently, finding explicit classical solutions -- such as those describing point-like electric and magnetic charges -- is particularly difficult. Recently, some exact plane-wave, Coulomb-type, and wave-packet solutions in $\kappa$-Minkowski space have been reported in \cite{kupriyanov2023lie}, \cite{abla2025poisson}, and \cite{Kurkov:2025soi}, respectively.

In this paper, we address these challenges by constructing explicit, rotationally invariant classical solutions for Maxwell--Chern--Simons theory on a three-dimensional noncommutative spacetime with a constant spacelike Poisson structure. These solutions provide a concrete realization of point-like charges in the noncommutative regime, offering new insights into the dynamics and regularization of charged objects in noncommutative gauge theories.

\section{Poisson gauge theory}\label{PGT}

We study a three-dimensional gauge theory defined on the Minkowski spacetime
$\mathbb{R}^{1,2}$ equipped with Cartesian coordinates $x^\mu=(x^0,x^1,x^2)$ and
a metric of signature $(-++)$. The spacetime is endowed with a Poisson structure
\begin{equation}
    \{x^\mu,x^\nu\}=\theta^{\mu\nu},
\end{equation}
where $\theta^{\mu\nu}$ is a constant antisymmetric tensor. In three
dimensions, $\theta^{\mu\nu}$ is necessarily degenerate and can be
parametrized as
\begin{equation}
    \theta^{\mu\nu}=\epsilon^{\mu\nu\lambda}\xi_\lambda
\end{equation}
for a given vector $\xi^\mu$. Depending on the causal nature of $\xi^\mu$,
the resulting noncommutativity is space-like, time-like, or light-like. In this
work, we focus on time-like noncommutativity, assuming
\begin{equation}\label{pb}
    \{x^0,x^i\}=0,\qquad
    \{x^i,x^j\}=g\,\epsilon^{ij},\qquad i,j=1,2 \,,
\end{equation}
where the parameter $g$ has the dimension of length squared and controls the scale of noncommutative deformation and $\epsilon^{ij}\epsilon_{kj} = \delta^i_k$.

The gauge field is a one-form $A=A_\mu(x)\,dx^\mu$ on $\mathbb{R}^{1,2}$. Infinitesimal gauge
transformations with parameter $\varepsilon(x)$ are postulated in the form  
\begin{equation}\label{gtr}
    \delta_\varepsilon A_\mu = D_\mu\varepsilon := \partial_\mu \varepsilon+\{A_\mu,\varepsilon\}\,.
\end{equation}
The commutator of two transformations closes according to
\begin{equation}
    [\delta_{\varepsilon_1},\delta_{\varepsilon_2}]
    =\delta_{\{\varepsilon_1,\varepsilon_2\}},
\end{equation}
implying that the underlying gauge algebra is non-Abelian. The field-strength tensor is defined as
\begin{equation}\label{F}
    F_{\mu\nu}=\partial_\mu A_\nu-\partial_\nu A_\mu+\{A_\mu,A_\nu\}\,,
\end{equation}
and satisfies the standard Bianchi identity
\begin{equation}\label{BI}
    D_{\lambda}F_{\mu\nu}+ D_{\mu}F_{\nu\lambda}+ D_{\nu}F_{\lambda\mu}=0
    \end{equation}
    with respect to the covariant derivative defined in (\ref{gtr}). Under the gauge transformations~\eqref{gtr} the field strength transforms in the adjoint
representation:
\begin{equation}
    \delta_\varepsilon F_{\mu\nu}=\{F_{\mu\nu},\varepsilon\}\,.
\end{equation}

Following the pattern of Yang--Mills theory, it is straightforward to define a Poisson analogue for Maxwell--Chern--Simons theory.
This is described by the action
\begin{equation}\label{MCS}
    {S_{MCS}=\int d^3x\left[
    -\frac{1}{4}F_{\mu\nu}F^{\mu\nu}
    +\frac\kappa2\epsilon^{\mu\nu\lambda}
    \left(A_\mu\partial_\nu A_\lambda+\frac13A_\mu\{A_\nu,A_\lambda\}\right)
    \right]\,,}
\end{equation}
which is gauge invariant up to a
boundary term.  Throughout this work, indices are raised and lowered using the Minkowski metric $\eta_{\mu\nu}$. By definition, the gauge field 
$A_\mu$ has the dimension of inverse length; consequently, both the Chern--Simons level $\kappa$ and the action $S_{MCS}$ share this dimension. 
 Varying the action yields the field equations
\begin{equation}
    D_\mu F^{\mu\nu} + \kappa F^\nu = 0,
    \qquad
    F^\nu:=\frac12\epsilon^{\nu\mu\lambda}F_{\mu\lambda}.
\end{equation}
The Maxwell term provides propagating degrees of freedom, while the
Chern--Simons term induces a topological mass \cite{SCHONFELD1981157, 1982AnPhy.140..372D}. 

The translation invariance of the action (\ref{MCS}) implies the conservation of energy and momentum. While the canonical energy-momentum tensor can be derived using Noether’s theorem, we present here a more geometric derivation. First, consider the symmetric tensor
\begin{equation}
    \Theta^{\mu\nu}=F^\mu F^\nu-\frac12\eta^{\mu\nu}F_\lambda F^\lambda\,.
\end{equation}
Computing its covariant divergence, we find
\begin{equation}
  D_\mu\Theta^{\mu\nu}=  F^\nu D_\mu F^\mu +F^\mu D_\mu   F^\nu -F_\lambda D^\nu F^\lambda\,.
\end{equation}
The first term on the right-hand side vanishes due to the Bianchi identity (\ref{BI}). The remaining two terms cancel on-shell, since 
\begin{equation}
    F^\mu D_\mu   F^\nu -F_\lambda D^\nu F^\lambda=F_\lambda( D^\lambda   F^\nu - D^\nu F^\lambda)\approx \kappa F_\lambda F_\mu\epsilon^{\lambda\mu\nu}
=0 \,.   \end{equation}
Therefore, we can write
\begin{equation}
    D_\mu\Theta^{\mu\nu}=\partial_\mu\Theta^{\mu\nu}+\{A_\mu,\Theta^{\mu\nu}\}=\partial_\mu T^{\mu\nu}\approx 0\,,
\end{equation}
where 
\begin{equation}
    T^{\mu\nu}=\Theta^{\mu\nu}-\theta^{\mu\alpha}\partial_\alpha A_\lambda \Theta^{\lambda\nu}
\end{equation}
is the desired energy-momentum tensor. Unlike in conventional gauge theory,  the tensor $T^{\mu\nu}$ is neither symmetric nor gauge invariant. Both properties are recovered in the commutative limit, where it reduces to the standard energy-momentum tensor of the electromagnetic field. 
Note that for our choice of Poisson brackets (\ref{pb}), the energy-momentum density $T^{0\mu}=\Theta^{0\mu}$ depends only on the field strength tensor $F_{\mu\nu}$, so that $ \delta_\varepsilon T^{0\mu}=\{\varepsilon, T^{0\mu}\}$. As a result, the total energy-momentum of the field 
\begin{equation}
    \mathcal{P}^\mu=\frac1\pi\int (F^0F^\mu-\frac12\eta^{0\mu}F^2)d^2x
\end{equation}
is a gauge-invariant vector quantity, provided that the gauge parameter $\varepsilon$ or the fields vanish at infinity. In particular, the total energy is given by the positive-definite  integral
\begin{equation}\label{en}
   \mathcal{E}= \mathcal{P}^0=\frac1{2\pi}\int( F^2_{01}+F^2_{02}+F^2_{12})d^2x\,.
\end{equation}

The asymmetry of $T^{\mu\nu}$ cannot be removed by a Belinfante--Rosenfeld shift. In the presence of the fixed background tensor 
$\theta^{\mu\nu}$, Lorentz invariance is explicitly broken. A symmetric energy-momentum tensor would generate conserved Lorentz currents, 
\begin{equation}\label{M}
M^{\mu\nu|\lambda}=x^\mu T^{\nu\lambda}-x^\nu T^{\mu\lambda}\,,\qquad \partial_\lambda M^{\mu\nu|\lambda}=0\,,
\end{equation}
contradicting this symmetry breaking. The asymmetry of 
$T^{\mu\nu}$ is therefore structural and inherent to the theory.

Nevertheless, there remains a residual Lorentz symmetry corresponding to the little group of the vector $\xi_\lambda=\epsilon_{\lambda\mu\nu}\theta^{\mu\nu}$. It is generated by the Hamiltonian vector field 
\begin{equation}
    R=\{H,-\}=x_\mu\theta^{\mu\nu}\partial_\nu\,,\qquad H:=\frac12\eta_{\mu\nu}
x^\mu x^\nu\,,
\end{equation}
and acts on $A_\mu$ through the Lie derivative:
\begin{equation}
    \mathcal{L}_RA_\mu=RA_\mu+\eta_{\mu\lambda}\theta^{\lambda\nu}A_\nu\,.
\end{equation}
Since the Lagrangian density $L$ of (\ref{MCS}) is constructed in terms of the $R$-invariant tensors  $\eta^{\mu\nu}$, $\epsilon^{\mu\nu\lambda}$, and $\theta^{\mu\nu}$,  we can write
\begin{equation}
    RL=\frac{\delta S}{\delta A_\mu}\mathcal{L}_RA_\mu + \partial_\nu \left(\frac{\partial L}{\partial (\partial_\nu A_\mu)}\mathcal{L}_RA_\mu\right)\,.
\end{equation}
The last equation is equivalent to the conservation law $\partial_\nu {M}^\nu\approx 0$, where 
\begin{equation}
    {M}^\nu=\frac{\partial L}{\partial (\partial_\nu A_\mu)}\mathcal{L}_RA_\mu+ \theta^{\nu\mu}x_\mu L\,.
    \end{equation}
The spatial integral for the total angular momentum of the field
\begin{equation}
    \mathcal{M}=\frac1\pi\int d^2x\, {M}^0
\end{equation}
simplifies considerably for the Poisson brackets (\ref{pb}). We obtain
\begin{equation}\label{m0}
    {M}^0=\Pi^i\mathcal{L}_R A_i\,,\qquad R=x_i\epsilon^{ij}\partial_j\,,
\end{equation}
where 
\begin{equation}
    \Pi^i=F^{i0}+{\frac{1}{2}}\kappa\epsilon^{ij}A_i
    \end{equation}
are the momenta canonically conjugate to the fields $A_i$.

Gauge symmetry also gives rise to a {\it lower-degree conservation law} \cite{anderson1996asymptotic,barnich2000local,barnich2002covariant, Sharapov:2016sgx, 10.21468/SciPostPhysLectNotes.77}. This stems from the global shift symmetry of  the gauge
transformations (\ref{gtr}),
\begin{equation}\label{shift}
\varepsilon\to \varepsilon +c\,,
\end{equation}  
where $c$ is an arbitrary constant. Applying the general procedure of Refs. \cite{barnich2002covariant}, \cite{Sharapov:2016sgx} yields an on-shell closed one-form $J=J_\mu dx^\mu$ constructed from the gauge field $A_\mu$ and its derivatives. In the case at hand, one can derive this one-form in a more direct way. The starting point is the Noether identity associated with the equations of motion:
\begin{equation}
    D_\nu D_\mu F^{\mu\nu}-\kappa D_\nu F^\nu\equiv  0\,.
\end{equation}
This is equivalent to $\partial_\nu B^\nu=0$, where the vector field
\begin{equation}
    B^\nu=(D_\mu F^{\mu\nu}-\kappa F^\nu) -\theta^{\nu\lambda}\partial_\lambda A_\rho(  D_\mu F^{\mu\rho}-\kappa  F^\rho)
\end{equation}
vanishes on shell. 
Since the divergence of  $B^\nu$ is identically  zero for arbitrary $A_\mu$, there  exists a bivector field $J^{\mu\nu}=-J^{\nu\mu}$ such that $B^\mu=\partial_\nu J^{\nu\mu}$. A direct verification confirms that the sum
\begin{equation}\label{cb}
        J^{\mu\nu}=J^{\mu\nu}_M+\kappa J^{\mu\nu}_{CS}\,,
\end{equation}
where
\begin{equation}\label{JM}
\begin{array}{rcl}
    J_M^{\mu\nu}&=& \displaystyle f^{\mu\nu} -   \frac{1}{4}\theta^{\mu\nu}f^{\alpha\beta}f_{\alpha\beta}- (\theta^{\alpha\mu}f^{\nu\beta} - \theta^{\alpha\nu}f^{\mu\beta})\partial_{\alpha}A_{\beta}\\[3mm]
         &+& \theta^{\mu\nu}A^{\beta}\partial^{\alpha}\{A_{\alpha},A_{\beta}\} + A^{\beta}\partial_{\alpha}(\theta^{\alpha\mu}\{A^{\nu},A_{\beta}\}- \theta^{\alpha\nu}\{A^{\mu},A_{\beta}\}) +\theta^{\mu\r}\theta^{\nu\lambda}\partial_{\r}A^{\alpha}\partial_{\lambda}A^{\beta}f_{\alpha\beta}\\[3mm]
         &+&\displaystyle \frac{1}{2}\Big(\theta^{\mu\nu}\big(A^{\beta}\{A^{\alpha},\{A_{\alpha},A_{\beta}\}\} + \frac{1}{2}\{A,A\}^2\big)
         + (\theta^{\mu\r}\theta^{\nu\lambda}- \theta^{\nu\r}\theta^{\mu\lambda})A^{\beta}\partial_{\r}A^{\alpha}\partial_{\lambda}\{A_{\alpha},A_{\beta}\}\Big)  
         \end{array}
         \end{equation}
         and
         \begin{equation}\label{JCS}
         \begin{array}{rcl}
    J_{CS}^{\mu\nu}&=&\displaystyle \varepsilon^{\nu\mu\alpha}A_{\alpha} + (\varepsilon^{\alpha\beta\nu}\theta^{\mu\lambda} - \varepsilon^{\alpha\beta\mu}\theta^{\nu\lambda})A_{\alpha}\partial_{\lambda}A_{\beta} + \frac{1}{2}\theta^{\mu\nu}A^{\alpha}f_{\alpha}\\[3mm]
         &+& \displaystyle \frac{1}{3}\varepsilon^{\alpha\beta\gamma}\big(\theta^{\mu\nu}A_{\alpha}\{ A_{\beta}, A_{\gamma} \} + 2\theta^{\mu\lambda}\theta^{\nu\rho}A_{\alpha}\partial_{\rho}A_{\beta}\partial_{\lambda}A_{\gamma}\big)  \,,
         \end{array}
\end{equation}
possesses the desired property.  Here, we have  employed the notation 
\begin{equation}
f_{\alpha} = 2\varepsilon_{\alpha\mu\nu}\partial^{\mu}A^{\nu}\,,\quad f_{\mu\nu} = \partial_{\mu}A_{\nu} - \partial_{\nu}A_{\mu}\,,\quad \{A,A\}^2 = \{A^{\mu},A^{\nu}\}\{A_{\mu},A_{\nu}\}\,.
\end{equation}
Upon dualization, we obtain a one-form $J=\epsilon_{\mu\nu\lambda}J^{\mu\nu}dx^\lambda$, which is closed on the equations of motion.
The relation $dJ\approx0$ represents a Poisson generalization of Gauss's law. Notice that the formula for $J_M$ is applicable in any dimension. Neither of the bivectors above survives in conventional Maxwell or Chern–Simons theory with a non-Abelian gauge group.  A lower-degree conservation law arises, however, in unimodular gravity \cite{elfimov2022lie}. 

There is a considerable ambiguity in the definition of conserved bivector (\ref{cb}): one may add terms that either vanish on-shell or can be expressed  as the divergence of a three-vector. However, this ambiguity does not affect the associated conserved charges. 
In the case of pure Chern--Simons theory, for example, the on-shell closed one-form
\begin{equation}\label{J}
    J=\big(A_\mu+\frac12\theta^{\nu\lambda}A_\nu\partial_\mu A_\lambda\big)dx^\mu
\end{equation}
is as valid as the one given by (\ref{JCS}). One readily  verifies that
\begin{equation}\label{dJ}
    dJ=\frac12F_{\mu\nu}\omega^\mu_\alpha \bar{v}^\nu_\beta dx^\alpha \wedge dx^\beta\approx 0\,,
\end{equation}
where 
\begin{equation}\label{wv}
    \omega^\mu_\alpha=\delta^\mu_\alpha+\theta^{\mu\nu}\partial_\alpha A_\nu\,,\qquad v^\nu_\beta=\delta^\nu_\beta+\partial_\mu A_\beta\theta^{\mu\nu} \,.
\end{equation}
For $\theta$ sufficiently small, the matrices $\omega$ and $v$ are invertible, and we denote $\bar v=v^{-1}$. The geometry of symplectic groupoids provides an alternative interpretation of this conservation law. Following \cite{kupriyanov2024symplectic}, the exact two-form $F^s=dJ$ represents the {\it gauge covariant} field strength associated with $A$. For a recent discussion regarding the relationship between different field-strength tensors, see 
\cite{di2026electromagnetic}. 

\section{Poisson--Chern--Simons theory} 
Let us start with a pure Chern--Simons theory. We seek $SO(2)$-invariant solutions of the flatness condition
\begin{equation}\label{F=0}
    F_{\mu\nu}=0\,.
\end{equation}
The most general rotation-invariant ansatz for $A$ is given by
\begin{equation}\label{notes_definition}
    A_0 = A_0(t, \rho)\,,\qquad A_{i} = x_{i}\widetilde{\Psi}(t,\rho) + \varepsilon_{ij}x^{j}\widetilde{\Phi}(t,\rho)\,,
\end{equation}
where $t=x^0$  and $\rho=\frac12(x_1^2+x_2^2)$; the functions $\widetilde{\Psi}$ and $\widetilde{\Phi}$ represent the polar and axial parts of the vector potential $A_i$. The gauge transformations (\ref{gtr}) for the $A_0$, $\widetilde{\Psi}$, and $\widetilde{\Phi}$ take the form
\begin{equation}\label{notes_gauge2}
    \delta_{\varepsilon}A_0 = \partial_t\varepsilon\,,\qquad \delta_{\varepsilon}\widetilde{\Psi} = (1 - g\widetilde{\Phi})\partial_{\rho}\varepsilon\,,\qquad \delta_{\varepsilon}\widetilde{\Phi} = g\widetilde{\Psi}\partial_{\rho}\varepsilon\,,
\end{equation}
where the gauge parameter $\varepsilon = \varepsilon(\rho,\,t)$.
Upon substituting (\ref{notes_definition}) into the definition (\ref{F}), we obtain the  system of equations 
\begin{equation}\label{zero_sys}
    \begin{array}{l}
     \displaystyle    F_{0i} = \frac{1}{g}\big(x_{i}(\partial_t\Psi - g\Phi\partial_{\rho}A_0) - \varepsilon_{ij}x^{j}(\partial_t\Phi + g\Psi\partial_{\rho}A_0)\big) = 0\,,\\[5mm]
   \displaystyle      F_{ij} = \frac{1}{g}\varepsilon_{ij}\partial_{\rho}\big(\rho(\Psi^2 + \Phi^2 - 1)\big) = 0\,,
    \end{array}
\end{equation}
where we introduced the notation 
\begin{equation}\label{phiPsi}
\Psi = g\widetilde{\Psi}\,,\qquad \Phi = 1 - g\widetilde{\Phi}\,.
\end{equation} 
Integration of the last equation yields
\begin{equation}
    \Psi^2 + \Phi^2 = 1 + \frac{c}{\rho}\,,
\end{equation}
where $c = c(t)$ is an arbitrary function of time. By parametrizing
\begin{equation}
     \Psi =  \displaystyle   \sqrt{1 + \frac{c}{\rho}}\,\sin{(g\varphi)}\,,\qquad
     \Phi =\displaystyle  \sqrt{1 + \frac{c}{\rho}}\,\cos{(g\varphi)}\,,
\end{equation}
for arbitrary function $\varphi(t,\rho)$, and substituting these into the remaining two equations
\begin{equation}
         \partial_t\Psi = g\Phi\partial_{\rho}A_0\,,\qquad \partial_t\Phi =- g\Psi\partial_{\rho}A_0\,,
\end{equation}
we find 
\begin{equation}
    \partial_t c=0\,,\qquad \partial_\rho A_0=\partial_t \varphi\,.
\end{equation}
Applying the gauge transformations (\ref{notes_gauge2}) shows that the fields $\varphi$ and $A_0$ are purely gauge:
\begin{equation}
    \delta_{\varepsilon}\varphi = \partial_{\rho}\varepsilon\,,\qquad \delta_{\varepsilon}A_0 = \partial_t\varepsilon\,.
\end{equation}
This allows us to impose the gauge fixing condition $A_0=\varphi=0$, leading to the solution
\begin{equation}\label{monopol}
 A_0=0\,,\qquad   A_{i} = \frac{1}{g}\varepsilon_{ij}x^{j}\left(1 - \sqrt{1 + \frac{c}{\rho}}\right)\,.
\end{equation}
All rotationally invariant solutions to the zero-curvature equation (\ref{F=0}) are therefore  gauge equivalent to stationary configurations labelled by a single constant $c$ with dimension length squared. Since the only fundamental parameter with this dimension is the coupling constant $g$, it is natural to set $c=-2qg$ for a dimensionless constant $q$. 

The solution exhibits an expected singularity at the origin $\rho=0$, which can be interpreted as a localized source. 
Indeed, in the commutative limit $g\to 0$, the solution (\ref{monopol}) reduces to a flat $U(1)$ connection
\begin{equation}\label{A0circ}
   A^\circ= \lim_{g\to 0} A =q\,\frac{x_1dx_2-x_2dx_1}{x_1^2+x_2^2}\,,
    \end{equation}
defined on the punctured plane $\mathbb{R}^2\backslash \{0\}$ and possessing the holonomy
\begin{equation}
 q=   \frac{1}{2\pi}\oint A^\circ\,.
\end{equation}
Furthermore, $A\to A^\circ$ as $\rho\to \infty$; hence, the one-forms (\ref{monopol}) and (\ref{A0circ}) coincide asymptotically. If $\gamma_\infty$ is a space-like circle of infinite radius, then the integral
\begin{equation}
    q=\frac{1}{2\pi}\oint_{\gamma_\infty}A
\end{equation}
represents the total charge of the field configuration.
Note that this is not the standard electric charge, since we are considering Chern--Simons theory rather than Maxwell theory. Instead, $q$ is naturally interpreted as a magnetic flux produced by a 2D analogue of an ideal Aharonov--Bohm solenoid localized at a single point. In conventional (2+1)-dimensional gauge theory, such magnetic fluxes are usually attributed to hypothetical point particles called {\it anyons} \cite{Anyons}. We adopt this terminology and henceforth refer to $q$ as the magnetic flux associated with a static anyon.

As discussed in Sec. \ref{PGT},  the Poisson--Chern--Simons theory admits a lower-degree conservation law (\ref{J}).  For the specific solution (\ref{monopol}), one readily finds  
$J=A^\circ$ and consequently
\begin{equation}
    q=\frac1{2\pi}\oint J\,,
\end{equation}
for any simple loop enclosing the origin. This provides a rigorous interpretation of (\ref{monopol}) as  the magnetic flux localized at a single point.

Note that the Poisson structure (\ref{pb}) explicitly  breaks invariance under spatial reflection
$x^i\to -x^i$. However, the model remains invariant under the combined transformation
\begin{equation}
    x^i\to -x^i\,,\qquad g\to -g\,,
\end{equation}
which simultaneously reverses the sign of the coupling constant. If the coupling
constant is restricted to be non-negative, $g\geq0$, this symmetry effectively corresponds to 
the reversal of the magnetic charge, $q\to -q$.

For $c=-2qg>0$, the solution (\ref{monopol}) is well-defined on the entire plane except
at the origin, $\rho=0$, where it is discontinuous.  Note that the discontinuity of the gauge potential $A_i(x)$ is finite and depends on the direction from which the singular point is approached.  In the commutative case, this limiting value is always infinite. Consequently, noncommutativity partially regularizes the singularities of anyonic solutions. 

For $c=-2qg<0$, additional singularities appear at $\rho=-c$.
In this case, the solution is well-defined only outside a disk of radius
$r=\sqrt{-2c}$ and becomes complex inside this region. Thus, positive and negative
fluxes (for a fixed sign of $g$) exhibit distinct analytic structures.
This asymmetry, however, is consistent with CPT invariance. 

From a physical viewpoint, there is nothing pathological about the breakdown of a classical
solution in a small spatial region. One should keep in mind that Poisson gauge theory
is, by construction, a low-energy effective approximation to a gauge theory on a genuinely
noncommutative spacetime. Accordingly, it is expected to be applicable only at length scales
much larger than the fundamental scale $\ell=\sqrt{g}$.

\section{Multianyon solutions}

The translation invariance of the equations of motion allows one to shift the solution (\ref{monopol}) into another stationary solution in which a static anyon is located at the point $y$:
\begin{equation}\label{mmonopol}
 A_0=0\,,\qquad   A_{i} = \frac{1}{g}\varepsilon_{ij}(x^{j}-y^j)\left(1 - \sqrt{1 + \frac{2qg}{(x_1-y_1)^2+(x_2-y_2)^2}}\right)\,.
\end{equation}
Starting from this three-parameter family of gauge potentials, we construct a more general solution that depends on $3n$ parameters, corresponding to a configuration of $n$ anyons. The nonlinearity of the field equations (\ref{F=0}) precludes a simple linear superposition of single-anyon solutions (\ref{mmonopol}). 

In \cite{kupriyanov2024symplectic}, it was suggested to interpret  the gauge field $A$ as a bisection of the symplectic groupoid 
over the Poisson manifold $\mathbb{R}^{1,2}$. 
The set of bisections forms a group $\mathcal{B}$, which acts on $\mathbb{R}^{1,2}$ via diffeomorphisms. The identity element of $\mathcal{B}$ is $A=0$, and the group multiplication is defined by
\begin{equation}\label{prod}
    A\circ A'= A_\mu\big(x^\nu+\theta^{\nu\lambda}A'_\lambda(x)\big)dx^\mu +A'_\mu(x)dx^\mu\,.
\end{equation}
For $\theta\neq 0$, this multiplication is both nonlinear and nonlocal. The action of a bisection $A$ on $\mathbb{R}^{1,2}$ is given by
\begin{equation}\label{act}
    x^\mu \mapsto  \phi_A^\mu(x)=x^\mu+\theta^{\mu\nu}A_\nu(x)\,,
\end{equation}
so that $\phi_A\circ\phi_{A'}=\phi_{A\circ A'}$. Bisections with vanishing curvature (\ref{F})
 are referred to as {\it Lagrangian}\footnote{The image of a Lagrangian bisection is a Lagrangian submanifold of the symplectic groupoid, hence the name.}. They constitute a subgroup $\mathcal{L}\subset \mathcal{B}$.
 Thus, the solution space of the field equations $F_{\mu\nu}=0$ is identified with the subgroup of Lagrangian bisections. The group multiplication (\ref{prod}) then provides a natural framework for the nonlinear superposition of particular solutions. 
 
Let $A(q,y)$ denote the single-anyon solution (\ref{mmonopol}). Given  $n$ such solutions with charges and locations $(q_k,y_k)$, we define the $n$-anyon configuration by
\begin{equation}\label{AA}
    A(q_1,y_1;\ldots; q_n,y_n)=A(q_1,y_1)\circ A(q_2, y_2)\circ \cdots \circ A(q_n,y_n)\,.
\end{equation}
Since the group $\mathcal{L}$ is non-Abelian, the resulting configuration depends on the ordering of the factors. 
Taken together, the various products (\ref{AA}) constitute a subgroup in $\mathcal{L}$. We denote this subgroup $\mathcal{A}$ and refer to as  the {\it anyon group}. 

The explicit substitution of (\ref{mmonopol}) into (\ref{AA}) leads to a rather cumbersome algebraic expression.  Therefore, we present only the two-anyon solution with one anyon located at the origin:
\begin{equation}\label{2an}
 A(q_1, 0;q_2,y)=a_1(z){\varepsilon_{ij}z^{j}dx^i} -a_2(z){\varepsilon_{ij}(z^{j}-y^{j})}dx^i\,,\end{equation}
where 
\begin{equation}
    a_1(z) = \frac1g \left(1-\sqrt{1-\frac{2q_1g}{z^2}}\right)\,,\qquad a_2(z) = \frac1g\left(1-\sqrt{1 + \frac{2q_2 g}{(z-y)^2}}\right)\,,
\end{equation}
and 
\begin{equation}\label{z}
    z^{i} = \phi^i_{A(q_2,y)}(x)=x^{i} - g(x^{i} - y^{i})a_2(x)\,,
\end{equation}
c.f. Eq.(\ref{act}). This solution simplifies considerably for $y=0$, when the two anyons are placed at the same point. In this case, we find that 
\begin{equation}
     A(q_1, 0;q_2,0)=A(q_1+q_2,0)\,,
     \end{equation}
 which is consistent with expectations: any rotation-invariant solution is defined by a single parameter -- the magnetic flux $q$ -- which is obviously additive in the commutative limit. In particular, superposing anyons with opposite fluxes results in $A(q,0;-q,0)=0$, which may be interpreted as anyon-antianyon annihilation. 
 In the opposite regime $y^2\gg g$, when the two anyons are well-separated, the interaction between them becomes negligible.   
Consequently, the system can be approximated as the linear superposition of two independent single-anyon solutions.

To further clarify the structure of the two-anyon solution (\ref{2an}), we compute the on-shell closed form (\ref{J}). After some algebraic manipulations, we obtain
\begin{equation}
    J=\frac{q_1\epsilon_{ij}z^jdz^i}{z^2}+\frac{q_2\epsilon_{ij}(z^{j} - y^{j})dz^i}{(z-y)^2}  -df\,, \end{equation}
where
\begin{equation}
    f= \frac{g}{2}(y_1z_2-y_2z_1)a_1(z)a_2(z)
\end{equation}
is a single-valued function that vanishes in the commutative limit. Integrating $J$ along infinitesimal loops allows one to locate the magnetic fluxes $q_1$ and $q_2$ at the points $z^i=0$ and $z^i=y^i$, respectively. We conclude that the interacting anyons preserve their magnetic fluxes, while their positions shift according to Eq. (\ref{z}).

The non-Abelian nature of the anyon group $\mathcal{A}$ is characterized by its  derived group $[\mathcal{A},\mathcal{A}]$. This is generated by the commutators 
\begin{equation}\label{com}
    [A_1,A_2]=A_1\circ A_2\circ A_1^{-1}\circ A_2^{-1}
\end{equation}
of multianyon solutions.  By construction, all elements of the derived group $[\mathcal{A},\mathcal{A}]$ vanish in the commutative limit. 
 For a pair of single-anyon solutions 
 $A_1(q_1,0)$ and $A_2(q_2,y)$, the commutator evaluates to
\begin{equation}\label{com1}
    [A_1,A_2]^{i}(x) = x^{i} + \Gamma^{i}_1\Big(y - \Gamma_2\big(y - \Gamma_{-1}(y + \Gamma_{-2}(x-y))\big)\Big)\,,
\end{equation}
where we have introduced the vector-valued functions
\begin{equation}
    \Gamma^{i}_{\pm a}(x) = x^{i}\sqrt{1\pm\frac{2q_a g}{x^2}}\,,\qquad a=1,2\,.
\end{equation}
To determine the positions of the anyons -- defined by the poles of the commutator \eqref{com} -- we utilize the general properties of the covariant field strength tensor \cite{kupriyanov2024symplectic}:
\begin{equation}
    F^s(A_1\circ A_2)=\phi^\ast_{A_2}F^s(A_1)+F^s(A_2)\,.
\end{equation}
Recalling the relation between the covariant strength tensor and the conserved current,  $F^s=dJ$, 
we obtain the composition law for the currents:
\begin{equation}
     J(A_1\circ A_2)=\phi^\ast_{A_2}J(A_1)+J(A_2)+df(A_1,A_2)\,,
\end{equation}
where $f$ is a scalar function. Consequently, if $y_1$ and $y_2$ denote the poles of $A_1$ and $A_2$, respectively, then the poles of the composition $A_1\circ A_2$ are located at $z_1=\phi_{A_2}(y_1)$ and $z_2=y_2$. Applying this result to the commutator (\ref{com}), we find four distinct poles at the positions:
\begin{equation}
    z_1=\phi^{-1}_{A_2\circ A_1\circ A_2^{-1}}(y_1)\,,\qquad z_2=\phi^{-1}_{A_2\circ A_1}(y_2)\,,\qquad z_3=\phi^{-1}_{A_2}(y_1)\,,\qquad z_4=y_2\,,
\end{equation}
carrying fluxes $q_1, q_2, -q_1$, and $-q_2$, respectively. So, the total magnetic flux is equal to zero. 

\section{Coulomb's potential}
We now consider the Poisson deformation of the source-free Maxwell's equations:
\begin{equation}\label{start}
    D^{\mu}F_{\mu\nu} = 0\,.
\end{equation}
Unlike the  Poisson--Chern--Simons theory discussed above, these equations describe propagating degrees of freedom. 
We are interested in solutions that would correspond to a deformation of the  Coulomb potential in conventional 3D electrodynamics. 
Therefore, we are looking for $SO(2)$-invariant stationary solutions. They are defined by the electromagnetic potentials (\ref{notes_definition}), where $\widetilde{\Phi}$ and $\widetilde{\Psi}$ are now functions of a single variable $\rho$. The components of the field strength tensor are given by Eq. (\ref{zero_sys}), where one should set $\partial_t\widetilde{\Phi}=\partial_t\widetilde{\Psi}=0$. Then the equations of motion take the form 
\begin{equation}\label{DF}
    \begin{array}{rcl}
         D^{\mu}F_{\mu0} &=& 2\partial_{\rho}\Big(\rho(\Psi^2 + \Phi^2)\partial_{\rho}A_0\Big) = 0\,,\\[5mm]
          D^{\mu}F_{\mu i} &=& \displaystyle \frac{1}{g}\Big[g^2(\partial_{\rho}A_0)^2 + 2\partial^2_{\rho}\big(\rho(\Phi^2 + \Psi^2)\big)\Big](x_{i}\Psi - \varepsilon_{ij}x^{j}\Phi)=0\,,
    \end{array}
\end{equation}
where we used the notation (\ref{phiPsi}). The residual gauge transformations (\ref{gtr}), compatible with the ansatz, are given by
\begin{equation}\label{gauge}
         \delta_{\varepsilon}A_0 = 0\,,\qquad
         \delta_{\varepsilon}A_{i} = -(\Phi x_{i} - \Psi\epsilon_{ij}x^j)\partial_{\rho}\varepsilon
\end{equation}
for $\varepsilon=\varepsilon(\rho)$. These transformations allow one to gauge away either of the two functions $\Phi$ and $\Psi$. 

As the second equation factorizes, the system (\ref{DF}) admits a simple solution: $\Psi=\Phi=0$ and $A_0(\rho)$ is arbitrary. Then 
\begin{equation}
    A_i=\frac{1}{g}\epsilon_{ij}x^j \quad \mbox{and}\quad F_{ij}=\frac1g\epsilon_{ij}\,.
\end{equation}
However, this solution lacks a well-defined commutative limit, and its physical interpretation remains obscure.

Substituting
\begin{equation}\label{sub}
    \Psi = R(\rho)\sin{\varphi(\rho)}\,,\qquad \Phi = R(\rho)\cos{\varphi(\rho)}\,,\qquad h(\rho) = \rho R^2(\rho)
\end{equation}
reduces the system  (\ref{DF}) to the following form:
\begin{equation}\label{eq}
    \begin{array}{l}
         \partial_{\rho}(h\partial_{\rho}A_0) = 0\,,\\[5mm]
         g^2(\partial_{\rho}A_0)^2 + 2\partial^2_{\rho}h = 0\,,
    \end{array}\quad\qquad\Longrightarrow\qquad\quad
    \begin{array}{l}
         h\partial_{\rho}A_0 = e\,,\\[3mm]
     \displaystyle    \partial^2_{\rho}h = -\frac{g^2e^2}{2h^2}\,,
    \end{array}
\end{equation}
where $e$ is an integration constant. By definition, $h(\rho)\geq 0$. Note that the function $\varphi(\rho)$ drops out of the equations due to gauge invariance (\ref{gauge}).  The last equation in (\ref{eq}) admits the first integral:
\begin{equation}\label{FI}
    \frac{1}{2}(\partial_{\rho}h)^2 -\frac{ g^2e^2}{2h} = \frac a2\,.
\end{equation}
Integrating this equation for the inverse function $\rho(h)$, we obtain
\begin{equation}\label{solution}
    \rho = \frac{h}{a}\sqrt{a + \frac{g^2e^2}{h}} - \frac{g^2e^2}{2a^{3/2}}\ln{\left(\frac{\sqrt{a + \frac{g^2e^2}{h}} + \sqrt{a}}{\sqrt{a + \frac{g^2e^2}{h}} - \sqrt{a}}\right)} - c\,.
\end{equation}

To provide a physical interpretation of the integration constants $e$, $a$, and $c$, we  consider the commutative limit of the solution. Since $g$ is the only parameter with the dimension of length squared, it is natural to parametrize $c$ as $c=q g$, where $q$ is a dimensionless constant. 
Then, in the commutative limit, the logarithmic term vanishes, and one obtains 
\begin{equation}\label{h=r}
    h=\sqrt{a}\rho\,.
\end{equation}
Combining Eqs.(\ref{eq}) and (\ref{FI}), we find
\begin{equation}\label{AM0}
    A_0 = \frac{e}{\sqrt{a}}\ln{\left(\frac{\sqrt{a+\frac{e^2g^2}{h}}+a}{\sqrt{a+\frac{e^2g^2}{h}} - a}\right)}=\frac e{\sqrt{a}}\ln h+ \mathrm{const}+\mathcal{O}\Big(\frac{e^2g^2}{h}\Big)\,.
\end{equation}
\begin{figure}
    \begin{minipage}[h]{0.49\linewidth}
        \center{\includegraphics[height = 6cm]{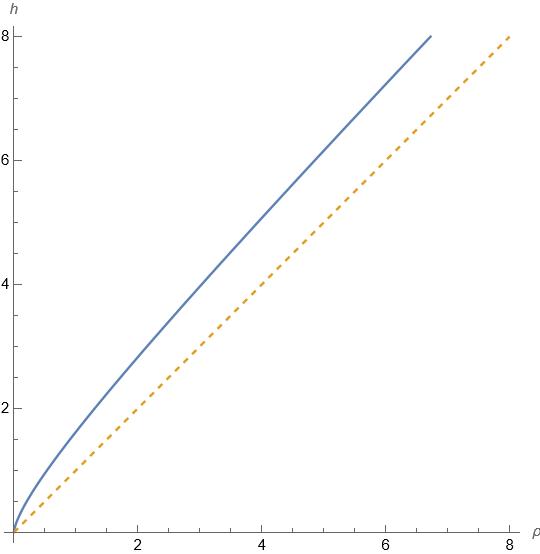} 
        }
    \end{minipage}
    \hfill
    \begin{minipage}[h]{0.49\linewidth}
        \center{\includegraphics[height = 6cm]{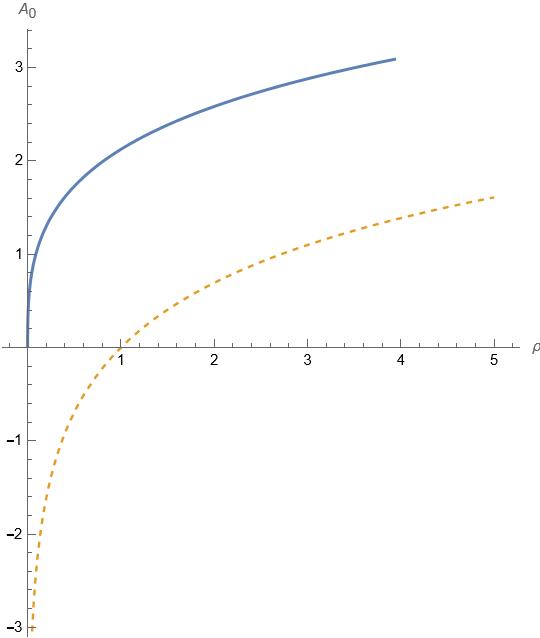} 
        }
    \end{minipage}
    \caption{ Left panel: Profile of the function $h(\r)$ for $c=0$, $q = 0$, $g=e =1$ (solid line). Right panel: Corresponding scalar potential $A_0(\r)$ (solid line); the dotted line shows the standard Coulomb potential ($g=0$) for comparison.}\label{Maxwellgraph}
    \end{figure}\\
Taking into account the asymptotic relation (\ref{h=r}), this reduces, in the commutative limit, to the standard Coulomb potential in $(2+1)$ dimensions:
\begin{equation}\label{stc}
    \lim_{g\to 0} A_0=\frac{e}{\sqrt{a}}\ln\rho\,,\qquad \lim_{g\to 0}A_i=0\,,
\end{equation}
where $k=1/\sqrt{a}$ plays the role of the electrostatic constant. In what follows, we set $a=1$.
 Relation (\ref{h=r}) also holds asymptotically in the limit $\rho\to \infty$ (see Fig. \ref{Maxwellgraph}).  Therefore, at large distances $\rho\gg g^2e^2$, the scalar potential $A_0(x)$ deviates only negligibly from its commutative counterpart (\ref{stc}). In the opposite regime $\rho\ll g^2e^2$, it follows from (\ref{solution}) that
\begin{equation}\label{00}
    h\sim \left[\frac{3ge
    }{2}(\rho+qg)\right]^{\frac23}\qquad \Longrightarrow\qquad A_0\sim\left[{\frac{12e}{g^2}(\rho +qg)}\right]^{\frac13}\,,
\end{equation}
which shows that the scalar potential remains finite at the origin, even for $q=0$. We emphasize that the detailed behavior of fields at length scales of order $\ell=\sqrt{g}$ has little physical significance  within the current approximation.

The presence of noncommutativity results in a nonzero  magnetic field. Substituting (\ref{sub}) into (\ref{zero_sys}), we find
\begin{equation}
    F_{ij}=-\frac1g\epsilon_{ij}\left(1-\sqrt{1+\frac{e^2g^2}{h}}\right)\sim\epsilon_{ij}\frac{g e^2}{2\rho}\quad \mbox{for}\quad \rho\gg g\,,
\end{equation}
so that the total magnetic flux through space diverges logarithmically. At the same time, the total energy of the magnetic field remains finite: 
\begin{equation}\label{E}
\begin{array}{c}
\displaystyle \mathcal{E}_{\mathrm{mag}}=\frac1{4\pi}\int_{\mathbb{R}^2} d^2x F_{ij}F^{ij}=\frac{1}{g^2}\int_0^\infty d\rho \left(1-\sqrt{1+\frac{e^2g^2}{h(\rho)}}\right)^2\\[7mm]
\displaystyle =\frac{1}{g^2}\int_0^\infty  \left(1-\sqrt{1+\frac{e^2g^2}{h}}\right)^2    \frac{dh}{\sqrt{1+\frac{e^2g^2}{h}}}= e^2 \,.
\end{array}
    \end{equation}
Remarkably, this energy does not vanish in the commutative limit. While gauge invariant, this quantity is of limited physical significance as it is not an integral of motion; only the total electromagnetic energy (\ref{en}) is conserved. The asymptotic behavior (\ref{h=r}) suggests that the total electrostatic energy diverges logarithmically, as in conventional 3D electrodynamics. Indeed, combining the expression for the field strength tensor (\ref{zero_sys}) with the equations of motion (\ref{eq}), we find that the electrostatic energy contained within the disk of radius $r$
is equal to
\begin{equation}\label{elen}
    \mathcal{E}_{\mathrm{el}}=\frac1{2\pi}\int_{D_r} (F_{01}^2+F_{02}^2)d^2x=e\big(A_0(r^2/2)-A_0(0)\big )\,.
\end{equation}
In view of (\ref{00}), the value $A_0(0)$ is finite, while $A_0(r^2/2)$ diverges logarithmically as $r\to\infty$.  Hence, the total electromagnetic energy is infinite, as expected\footnote{The energy (\ref{elen}) can be interpreted as the work required to move a point charge $e$, placed in the electrostatic potential $A_0$, from the origin to the boundary of the disk. For the conventional Coulomb  potential (\ref{stc}), the analogous integral (\ref{elen}) over space diverges at both the  lower and upper limits.}. 

The rotation invariance of the solution implies that $\mathcal{L}_R A=0$  and consequently, the angular momentum density \eqref{m0} also vanishes.

Eq. (\ref{E}) demonstrates that the constant $e$ possesses an invariant meaning and is not merely a gauge-fixing artifact.  An alternative way to relate this constant to the electric charge is through the nonlinear Gauss law (\ref{JM}). The total electric charge $Q$ of a given field configuration is obtained by integrating the dual of the bivector $J^{\mu\nu}_M$ over a circle at spatial infinity. The asymptotic behavior (\ref{stc}) implies that only the first term in (\ref{JM}) contributes to the integral as the radius $r$ of the circle goes to infinity. This yields
\begin{equation}
    Q=\lim_{r\to\infty}\frac1{8\pi}\oint \epsilon_{\mu\nu\lambda}J_M^{\mu\nu}dx^\lambda=\lim_{r\to\infty}\frac1{8\pi}\oint \epsilon_{\mu\nu\lambda}f^{\mu\nu}dx^\lambda=\lim_{r\to\infty}\frac1{4\pi}\oint \epsilon_{ij}\partial^iA_0dx^j=  e\,.  
\end{equation}
 It is also instructive to set $q=0$ and consider the opposite limit $r\to 0$, when the circle shrinks to the origin. In this case,  only a single term in the second line of Eq.(\ref{JM}) contributes to the integral, leading to
\begin{equation}
    \begin{array}{rcl}
          Q &=&\displaystyle \lim\limits_{r\rightarrow0}\frac1{8\pi}\oint \epsilon_{\mu\nu\lambda}J_M^{\mu\nu}dx^\lambda=\lim\limits_{r\rightarrow0}\frac1{8\pi}\oint\epsilon_{\mu\nu\lambda}\big(A^{\beta}\partial_{\alpha}(\theta^{\alpha\mu}\{A^{\nu},A_{\beta}\}- \theta^{\alpha\nu}\{A^{\mu},A_{\beta}\})\big)dx^{\lambda}\\[3mm]
          &=&\displaystyle -\lim\limits_{r\rightarrow0}\frac{1}{4\pi}\oint g^2\big(\epsilon_{ij}x^j(\w{\Psi}^2+\w{\Phi}^2) + 2\r x_i(\w{\Psi}\p_{\r}\w{\Phi} - \w{\Phi}\p_{\r}\w{\Psi})\big)\p_{\r}A_0dx^i\\[3mm]
          &=&\displaystyle \lim\limits_{r\rightarrow0}\frac{1}{2\pi}\int\limits_0^{2\pi}\big(g^2\r(\w{\Psi}^2 + \w{\Phi}^2)\p_{\r}A_0\big)d\phi = \lim\limits_{\r\rightarrow0}(g^2\r(\w{\Psi}^2 + \w{\Phi}^2)\p_{\r}A_0) = e\,,
    \end{array}
\end{equation}
where $x_1 = \sqrt{2\r}\cos{\phi}$, $x_2 = \sqrt{2\r}\sin{\phi}$. This demonstrates that the entire  electric charge is localized at the  point $r=0$. 

Turning off the electric charge ($e=0$) recovers the anyonic solution (\ref{monopol}) with flux $q$. This behavior is expected since all solutions to the zero-curvature condition (\ref{F=0}) identically satisfy the Maxwell equations (\ref{start}). Thus, a non-vanishing flux 
$q$ indicates a nontrivial superposition of the anyonic configuration and the genuine Coulomb potential.

It is noteworthy that the solutions (\ref{solution}) and (\ref{AM0}), when viewed as functions of the deformation parameter $g$, are non-analytic at $g=0$ and therefore cannot be captured by perturbative expansions about this point. Nevertheless, analyticity can be effectively restored by an appropriate shift of the integration constants, which is equivalent to the following renormalization of the magnetic flux and the scalar potential:
\begin{equation}
q \;\to\; q + \frac{g e^2}{a^{3/2}} \ln g \,,
\qquad
A_0 \;\to \;A_0 + \frac{2e}{\sqrt{a}} \ln g \,.
\end{equation}
The emergence of such nonanalytic behavior in the deformation parameter $g$ casts some doubt upon perturbative computations for the noncommutative Coulomb potential (see, e.g., \cite{gaete2004coulomb}, \cite{stern2008noncommutative}).

\section{Poisson--Maxwell--Chern--Simons theory}
We finally turn to a Coulomb-like solution in the full Poisson--Maxwell--Chern--Simons theory, governed by the field equations:
\begin{equation}\label{MaxChern}
    D^{\mu}F_{\mu\nu} + \frac\kappa2\varepsilon_{\nu\alpha\beta}F^{\alpha\beta} = 0\,.
\end{equation}
Using the same stationary, rotation-invariant ansatz as before, we reduce these equations to the following system:
\begin{equation}\label{Ah}
        \partial_{\rho}A_0 = \frac{2ge+\kappa \rho}{2gh} - \frac{\kappa}{2g}\,,\qquad
        \left(\frac{2ge+\kappa\rho}{2h}\right)^2 + 2\partial^2_{\rho}h = \frac{\kappa^2}{4}\,,
\end{equation}
where $e$ is an integration constant. For $\kappa=0$, these equations recover (\ref{eq}). Unlike the pure Maxwell case, the second equation admits no obvious first integral, which complicates  exact integration. Since the free (commutative) theory is known to describe a massive particle, the scalar potential $A_0$ produced by a point source is expected to be short-range. This implies the following asymptotic behavior:
\begin{equation}
    A_0\to 0\,,\qquad \partial_\rho A_0\to 0\quad \mbox{as}\quad \rho\to\infty\,.
\end{equation}
The system (\ref{Ah}) possesses the following exact solution satisfying the asymptotic conditions: 
\begin{equation}\label{an0}
    h=\rho+\frac{2ge}{\kappa}\,,\qquad A_0=0\,.
\end{equation}
This solution corresponds to a purely anyonic configuration: 
\begin{equation}\label{an}
    A_i= \frac{1}{g}\varepsilon_{ij}x^j\left(1 - \sqrt{1 + \frac{2ge}{\kappa \r}}\right)\,,
\end{equation}
with the flux $q=-e/\kappa$, cf. (\ref{monopol}).

Using the constant $\kappa$, it is convenient to introduce the  dimensionless variables 
\begin{equation}
    \bar{h}=\frac{\kappa^2}{4}h\,,\qquad \bar\rho=\frac{\kappa^2}{4}\rho+\bar{\rho}_0\,,\qquad \bar{\rho}_0=\frac{\kappa ge}{2}\,.
\end{equation}
In terms of these variables, Eq. (\ref{Ah}) takes the form:
\begin{equation}\label{hh}
    2\frac{d^2\bar{h}}{d\bar\r^2}+\left(\frac{\bar\r}{\bar{h}}\right)^2=1\,.
\end{equation}
The solutions to the equations are connected by the relation
\begin{equation}
    h(\rho)=\frac4{\kappa^2}\bar{h}\Big(\frac{\kappa^2}{4}\rho+\frac{\kappa ge}{2}\Big)\,.
\end{equation}
The form of the left hand side of Eq.(\ref{hh}) suggests two possible asymptotics for solutions as $\bar{\r}\to\infty$: linear $\bar{h}=\bar\r$ and quadratic $\bar{h}=\bar\r^2/2$. The latter leads to a growing scalar potential:
\begin{equation}
    \partial_\rho A_0=-\frac{\kappa}{2g}\quad \Longrightarrow\quad A_0\sim \rho\,.
\end{equation}
This does not correspond to a short-range potential. Therefore, we focus on solutions with linear asymptotics and write
\begin{equation}\label{hy}
    \bar{h}=\bar\r+Y(\bar\r)
\end{equation}
for some function $Y(\bar\r)$ that vanishes at infinity. Such solutions can be viewed as perturbations of the exact anionic solution (\ref{an}).  Upon substituting into (\ref{hh}), we find
\begin{equation}\label{yy}
    \left(\frac{\bar\r}{\bar\r+ Y}\right)^2 +2\frac{d^2 Y}{d\bar\r^2}=1\,.
\end{equation}
To extract the asymptotic behavior, we approximate the equation for large $\bar \r$, where nonlinear corrections become subleading:
\begin{equation}
    \frac{d^2Y}{\partial \bar\r^2}= \frac{Y}{\bar\r}\,.
\end{equation}
The decaying solution to this equation is expressed in terms of the modified Bessel function of the second kind:
\begin{equation}
    Y(\bar\r)=C \sqrt{\bar\r}\,K_1\Big(2\sqrt{\bar\r}\Big) \sim C \frac{\sqrt{\pi}}{2}{\sqrt[4]{\bar\r}}\; e^{-2\sqrt{\bar\r}} \,.
\end{equation}
Using the identity $K_1=-K_0'$ for the Bessel functions, we can integrate the first equation in (\ref{Ah}) within the same approximation:
\begin{equation}\label{A0}
    \partial_\rho A_0=-\frac{\kappa}{2\rho}Y\quad \Longrightarrow\quad A_0={\frac{1}{2}}C \kappa K_0(\kappa\sqrt{\rho})\,.
\end{equation}
In the asymptotic region, this gives the two-dimensional Yukawa potential 
\begin{equation}\label{Y}
    A_0\sim {\frac{1}{\sqrt[4]{\rho}}}\;e^{-\kappa\sqrt{\rho}}\,,
    \end{equation}
where $\kappa$ can be interpreted as the mass of the gauge boson. We note that the approximate scalar potential (\ref{A0}) is in fact an exact
solution of the Helmholtz--Yukawa equation in two dimensions: 
\begin{equation}
    (\nabla^2-\kappa^2)A_0(x)=-Q\delta^2(x)\,,\qquad Q=2\pi \kappa C\,.
\end{equation}
Therefore, at large distances, our solution coincides with the scalar potential generated by a point electric charge $Q$ in 3D massive electrodynamics.  
\begin{figure}
    \begin{minipage}[h]{0.49\linewidth}
        \center{\includegraphics[height = 6cm]{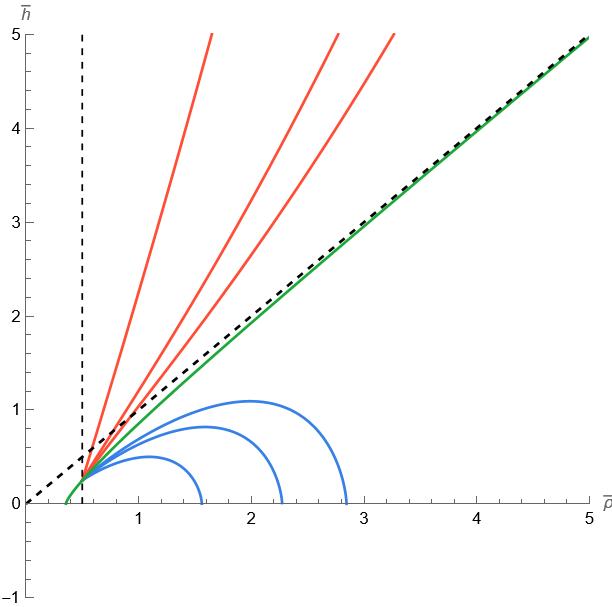} 
        }
    \end{minipage}
    \hfill
    \begin{minipage}[h]{0.49\linewidth}
        \center{\includegraphics[height = 6cm]{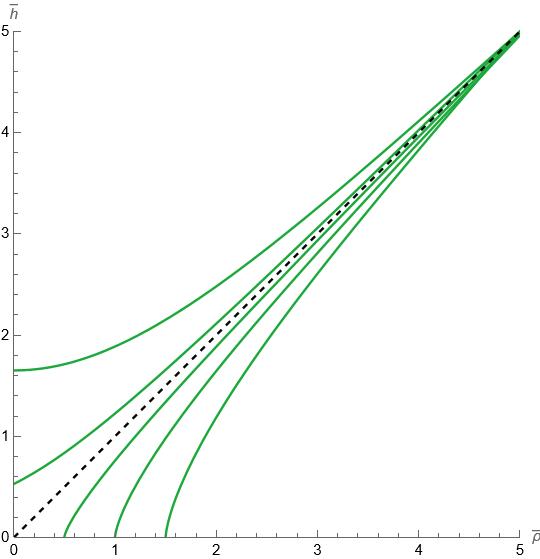} 
        }
    \end{minipage}
\caption{Left panel: Representative solutions of Eq.~(\ref{hh}) with the initial condition $\bar{h}(0.5)=0.25$. The green curve separates solutions with quadratic asymptotic behavior (red curves) from those that collapse at finite $\bar{\r}$ (blue curves). Right panel: Solutions with linear asymptotics; all cross the line $\bar{h}=0$.}
   \label{MCSgraph}\end{figure}
    
Numerical integration reveals that the second-order differential equation (\ref{hh}) admits a unique solution with linear asymptotics, $\bar{h}\sim \bar\r $, and satisfying the initial condition $\bar{h}(\bar\r_0)=0$ (see Fig. \ref{MCSgraph}). Equivalently, for any prescribed initial value $\bar{h}_0=\bar{h}(\bar\r_0)$, there exists a unique  initial slope $\bar{h}'(\bar\r_0)$ that ensures linear growth at large $\bar\r$. Each such integral curve acts as a separatrix between two distinct classes of solutions: those exhibiting quadratic growth and those that collapse to zero at a finite value of $\r$.   This yields   a one-parameter family of Yukawa-type solutions. To clarify the physical meaning of the parameter $\bar{h}_0$, we compute the magnetic flux passing out of the two-dimensional space:
\begin{equation}
\begin{array}{c}
  \displaystyle  \mathcal{H}=\frac1{8\pi}\int F_{12}\,d^2x = \frac{1}{2g}\int_0^\infty (\partial_\r h-1)\,d\r=\frac{2}{g\kappa^2}\int_{Y(\bar\r_0)}^\infty dY\\[5mm]
  \displaystyle =-\frac{2}{g\kappa^2}Y(\bar\r_0)=\frac{e}{\kappa}- \frac{2\bar{h}_0}{g\kappa^2}\,.
    \end{array}
\end{equation}
The first term on the right-hand side is independent of $g$ and corresponds to the magnetic flux generated by a point electric charge $e$ in conventional Maxwell--Chern--Simons theory \cite{SCHONFELD1981157, 1982AnPhy.140..372D}.
Consequently, a nonzero value of $\bar{h}_0$ can be viewed as signaling a partial mixing of the anyonic solution with a genuine Yukawa-type contribution.   Notably, in the pure anyonic case (\ref{an}), the magnetic flux vanishes. 

To analyze the behavior of the solution  $\bar{h}(\bar\r)$ near $\bar\r_0$, we approximate the differential equation (\ref{hh}) as follows:
\begin{equation}\label{hhh}
    2\frac{d^2\bar{h}}{d\bar\r^2}+\left(\frac{\bar\r_0}{\bar{h}}\right)^2=1\,.
\end{equation}
This equation admits a first integral
\begin{equation}
    \left(\frac{d\bar{h}}{d\bar\r}\right)^2-\frac{\bar\r_0^2}{\bar{h}}-\bar{h}=c  \,,
    \end{equation}
which yields the implicit solution 
\begin{equation}
    \bar\r=\int \frac{\sqrt{\bar h} \, d\bar h}{\sqrt{\bar h^2+c\bar h+\bar\r_0^2}}\,.
\end{equation}
Although this elliptic integral cannot be expressed in terms of elementary functions, it allows for a detailed examination of the special case
 $\bar h_0=0$. For this initial condition, approximating the denominator by setting  $\bar h\to 0$ yields the following asymptotic behavior: 
\begin{equation}\label{2/3}
    \bar h \sim\left(\frac{3}{2}\bar \r_0(\bar \r-\bar \r_0)\right)^{\frac23}\,.
\end{equation}
This behavior is analogous to that of the function $h$ in pure Maxwell theory, as derived in Eq. (\ref{00}).

We now evaluate the total electromagnetic energy of the Yukawa-type solutions above. Substituting in Eq.(\ref{en}), we obtain 
\begin{equation}
    \mathcal{E}=\mathcal{E}_{\mathrm{el}}+\mathcal{E}_{\mathrm{mag}}\,,
\end{equation}
where the electrostatic and magnetic contributions are given by
\begin{equation}
    \mathcal{E}_{\mathrm{el}}=\frac1{2\pi}\int (F_{01}^2+F_{02}^2) d^2x=2\int_0^\infty h(\partial_\r A_0)^2d\r=\frac{8}{g^2\kappa^2}\int_{\bar\r_0}^\infty \frac{Y^2(\bar \r)}{\bar{h}(\bar\r)}d\bar\r
\end{equation}
and
\begin{equation}
    \mathcal{E}_{\mathrm{mag}}=\frac1{2\pi}\int F_{12}^2 \,d^2x=\frac{8}{g^2\kappa^2}\int_{\bar \r_0}^\infty\left(\frac{d\bar h}{d\bar \r }-1\right)^2d\bar\r=\frac{8}{g^2\kappa^2}\int^\infty_{\bar\r_0}\left(\frac{dY}{d\bar\r}\right)^2 d\bar \r\,.
\end{equation}
For $\bar h_0 > 0$, both integrals converge since their integrands are continuous and rapidly decaying at infinity. In the special case $\bar h_0=0$, Eq. (\ref{2/3}) implies that the singular behavior near the lower integration limit is integrable. Hence, the entire family of Yukawa-type potentials is characterized by finite  electromagnetic  energy.

Finally, we apply the nonlinear Gauss law (\ref{cb}) to compute the total electric charge of the Yukawa-type solution above. As the integration contour, we choose a space-like circle at spatial  infinity. Due to the short-range behavior of the scalar potential $A_0$, the Maxwell bivector (\ref{JM}) does not contribute to the electric charge. At large distances from the origin, the solution approaches the exact anyonic configuration (\ref{an}). Therefore, the only non-vanishing contribution comes from the first term in the Chern--Simons bivector (\ref{JCS}), yielding:
\begin{equation}
\begin{array}{rcl}
    Q&=&\displaystyle\frac{1}{8\pi}\lim_{r\rightarrow\infty}\oint\epsilon_{\mu\nu\lambda}J^{\mu\nu}dx^{\lambda} = \frac{\kappa}{4\pi}\lim_{r\rightarrow\infty}\oint\epsilon_{ij}\,J^{0i}_{CS}\,dx^{j}\\[5mm]
     &=&\displaystyle-\frac{\kappa}{4\pi}\lim_{r\rightarrow\infty}\oint A_{i}dx^i = -\frac{1}{4\pi}\lim_{r\rightarrow\infty}\oint\frac{e}{\r}\epsilon_{ij}x^idx^j = e\,.
    \end{array}
\end{equation}
This confirms the interpretation of the constant $e$
 as the total electric charge of the configuration.

\section{Conclusion}

In this paper, we have constructed rotationally invariant solutions to the Maxwell--Chern--Simons equations in a noncommutative spacetime. Our analysis demonstrates that at large distances ($r \gg \sqrt{g}$), these solutions asymptotically recover their commutative counterparts, ensuring the correspondence principle at macroscopic scales. However, in the UV limit ($r \to 0$), the noncommutative geometry acts as a natural regulator, either resolving the central singularities or fundamentally altering the short-range behavior. Most notably, we established that the total electromagnetic energy of the Yukawa-type potential becomes finite, thereby resolving  the ``electromagnetic mass'' problem.

Furthermore, the non-analytic dependence of some solutions on the noncommutativity parameter $g$ suggests that standard perturbative expansions may fail to capture essential features of the theory. In the pure Chern--Simons case, we have generalized the multianyon solutions and revealed that their nonlinear superposition is governed by a Lie groupoid. This endows the solution space $\mathcal{A}$ with a non-Abelian group structure, potentially impacting the braiding statistics of these objects. Finally, the derived generalized Gauss law provides a robust framework for interpreting the physical parameters of the exact solutions.

It would be of great interest to extend these results to higher dimensions and to investigate other types of noncommutativity within this framework.

\paragraph{Acknowledgments.} This research was funded by the Ministry of Science and Higher Education of the Russian Federation under the research project FSWM-2025-0007. DSh gratefully acknowledges the support of the Foundation for the Advancement of Theoretical Physics and Mathematics ``BASIS''.

\end{document}